\documentclass{article}
\usepackage{amsmath}
\usepackage[dvips]{graphics}
\usepackage{latexsym}

\begin{document}
\newcommand{\bea}{\begin{eqnarray*}}
\newcommand{\eea}{\end{eqnarray*}}
\newcommand{\bean}{\begin{eqnarray}}
\newcommand{\eean}{\end{eqnarray}}
\newcommand{\eqs}[1]{Eqs. (\ref{#1})}
\newcommand{\eq}[1]{Eq. (\ref{#1})}
\newcommand{\meq}[1]{(\ref{#1})}
\newcommand{\fig}[1]{Fig. \ref{#1}}

\newcommand{\tri}{\delta}
\newcommand{\grad}{\nabla}
\newcommand{\pa}{\partial}
\newcommand{\pf}[2]{\frac{\pa #1}{\pa #2}}
\newcommand{\cla}{{\cal A}}
\newcommand{\aqt}{\frac{1}{4}\theta}

\newcommand{\om}{\omega}
\newcommand{\omo}{\omega_0}
\newcommand{\mo}{ M_{\rm o}}
\newcommand{\mi}{ M_{\rm i}}
\newcommand{\qo}{ Q_{\rm o}}
\newcommand{\qi}{ Q_{\rm i}}
\newcommand{\rop}{r_{{\rm o}+}}
\newcommand{\rom}{r_{{\rm o}-}}
\newcommand{\rip}{r_{{\rm i}+}}
\newcommand{\rim}{r_{{\rm i}-}}

\newcommand{\oh}{\frac{1}{2}}
\newcommand{\hsp}{\hspace{0.1mm}}
\newcommand{\spa}{\hspace{3mm}}
\newcommand{\hst}{\ \ \ }
\newcommand{\upd}[2]{^#1\hsp_#2}
\newcommand{\ri}{R_i}
\newcommand{\ro}{R_o}
\newcommand{\eqn}{&=&}
\newcommand{\non}{\nonumber \\}
\newcommand{\ppa}[2]{\left(\frac{\partial}{\partial #1}\right)^{#2}}
\newcommand{\pp}[2]{\frac{\partial #1}{\partial #2}}
\newcommand{\rn}{Reissner-Nordstr\"om}
\newcommand{\vphi}{\varphi}
\newcommand{\kss}{\left[K_\theta^\theta \right]}
\newcommand{\ktt}{\left[K_\tau^\tau \right]}
\newcommand{\heq}{{\hat =}}

\title{\bf Stability of Brans-Dicke thin shell wormholes}
\author{Xiaojun Yue\footnote{Email: yuexiaojun@mail.bnu.edu.cn} and Sijie Gao\footnote{Corresponding author.  Email: sijie@bnu.edu.cn} \\
Department of Physics, Beijing Normal University,\\
Beijing 100875, China}
\maketitle

\begin{abstract}
Recently, a class of spherically symmetric thin-shell wormholes in Brans-Dicke gravity have been introduced. Such wormholes can be supported by matter satisfying the weak energy condition (WEC). In this paper, we first obtain all the exact solutions satisfying the WEC. Then we show these solutions can be stable for certain parameters. A general requirement for stability is that $\beta^2>1$, which may imply that the speed of sound exceeds the speed of light. \\ \\
keywords: thin shell wormhole, Brans-Dicke gravity, stability
\end{abstract}

\section{Introduction}
After the pioneering work by Morris and Thorne \cite{morris}, many wormhole solutions have been found and analyzed. Visser \cite{visa1} proposed a simple method to construct thin shell wormholes by using Israel's junction condition \cite{israel66}. Following Visser's prescription, various thin shell wormholes have been discussed \cite{tw1}-\cite{lemos}.

It is well known that traversable wormholes within the framework of general relativity always violate energy conditions (see e.g.\cite{visae} and references therein). Since it is impossible to find wormholes with normal matter in Einstein's theory,  it is natural to look for wormhole solutions in alternative theories of gravity. A successful model was introduced by Mazharimousavi, {\em et.al.} \cite{bonnet}, where a thin-shell wormhole was constructed in five-dimensional Einstein-Mazwell-Gauss-Bonnet gravity. This wormhole is supported by normal matter and is stable against linear perturbation. Recently,  Eiroa {\em et.al.} \cite{bd2009} have proposed a thin-shell wormhole in four-dimensional Brans-Dicke gravity. The crucial feature of the wormhole is that it is supported by matter satisfying the weak energy condition (WEC). It is then important to know whether such wormholes are stable. In this paper, we show that stable Brans-Dicke wormholes with matter satisfying the WEC exist for a wide range of parameters. However, all solutions must satisfy $\beta^2>1$. The physical meaning will be discussed in the Conclusions.

This paper is organized as follows. In section \ref{deri}, we review the Brans-Dicke thin shell solutions and derive the conditions under which the wormhole is stable. In section \ref{cons}, we list all the necessary constraints for the existence of a Brans-Dicke wormhole satisfying the WEC and find all possible solutions for $\omega<-2$. We prove in Appendix \ref{app}  that no solutions can be found for $\omega>-3/2$. In section \ref{solution} ,  we find all stable solutions satisfying the WEC. We also prove that stable solution cannot exist if $\beta^2<1.866$. Finally, conclusions are made in section \ref{conc}.

\section{Derivation of the stability conditions} \label{deri}
Stability of thin-shell wormholes has been discussed in detail in the context of general relativity (see e.g. \cite{tw1,lobo, tw2}). We shall extend these methods to static Brans-Dicke wormholes.  Although our purpose is to investigate the static wormhole introduced in \cite{bd2009}, it is necessary to start with a dynamical wormhole since we need  perturb the wormhole  to test its stability.
The spacetime metric is assumed in the form
\bean
ds^2=-f(r)dt^2+g(r)dr^2+h(r)(d\theta^2+\sin^2\theta d\vphi^2)\,.
\eean
Suppose that the throat of the wormhole moves along the trajectory
\bean
r=r(\tau)\,,
\eean
where $\tau$ is the proper time. Let $\Sigma$ denote the hypersurface
$r=r(\tau)$. The matter is concentrated at the throat, while the rest of
the spacetime is the vacuum Brans-Dicke solution given by
\cite{R11}
\bean
f(r)\eqn \left(1-\frac{2\eta}{r}\right)^A \\
g(r)\eqn \left(1-\frac{2\eta}{r}\right)^B \\
h(r)\eqn \left(1-\frac{2\eta}{r}\right)^{1+B}r^2 \\
\phi(r)\eqn \phi_0\left(1-\frac{2\eta}{r}\right)^{-(A+B)/2} \,.
\eean
with
\bean
A=\frac{1}{\lambda}\,, \spa B=-\frac{\zeta+1}{\lambda},\spa
\lambda=\sqrt{(\zeta+1)^2-\zeta\left(1-\frac{\omega\zeta}{2}\right)}, \spa
\phi_0=\frac{4+2\omega}{3+2\omega}   \,,\label{laxi}
\eean
where $\eta$ and $\zeta$ are constants.

The unit tangent to $\Sigma$ in the
radial direction is
\bean
u^a=\dot t \ppa{t}{a}+\dot r \ppa{r}{a} \,,
\eean
where
\bean
\dot t(\tau) =\frac{1}{\sqrt{f(a)}}\sqrt{1+g(a)\dot r^2}\,, \label{tfa}
\eean
such that $g_{ab}u^au^{b}=-1$ is satisfied. The norm to $\Sigma$ can be written as
\bean
n_a=-\sqrt{g(r)f(r)}\dot r dt_a+\sqrt{g(r)(1+g(r)\dot r^2)}dr_a \,.
\eean
Let $\tau, \theta,\vphi$ be the coordinates of
$\Sigma$. Then the coordinate transformation is given by
\bean
t\eqn t(\tau) \\
r\eqn r(\tau) \\
\theta\eqn \theta \\
\vphi \eqn \vphi \,.
\eean
With respect to $\tau, \theta, \varphi$, the induced 3-metric on $\Sigma$ is written as
\bean
ds^2=-d\tau^2+h[r(\tau)](d\theta^2+\sin^2\theta d\varphi^2) \,. \label{3me}
\eean

The matter is located at the throat $\Sigma$ and described by the surface stress tensor $S_{ab}$.
Across the thin-shell throat, the junction conditions in Brans-Dicke theory take the form
\cite{bd2009}
\bean
-[K^i_j]+[K]\delta^i_j\eqn
\frac{8\pi}{\phi}\left(S^i_j-\frac{S}{3+2\omega}\delta^i_j\right) \,,
\label{j1}
\\
\left[\partial_n\phi \right] \eqn\frac{8\pi S}{3+2\omega} \,.
\label{j2}
\eean
where
\bean
[K^i_j]\equiv K^{i+}_j-K^{i-}_j
\eean
is the jump of $K^i_j$ across $\Sigma$.
 \eq{j1} is equivalent to
\bean
S^i_j=\frac{\phi}{8\pi}\left(\frac{\omega+1}{\omega}[K]\delta^i_j-[K^i_j]\right)\,.
\label{j3}
\eean

According to \cite{tw2}, the components of $K_{ab}$ can be written in the form
\bean
K_{ij}=-n_\gamma\left(\frac{\partial^2 x^\gamma}{\partial
\xi^i\partial
\xi^j}+\Gamma^\gamma_{\alpha\beta}\pp{x^\alpha}{\xi^i}\pp{x^\beta}{\xi^j}\right)
\,,
\label{kijn}
\eean
where $\{x^\alpha\}$ represent the spacetime coordinates and $\{\xi^i\}$ represent the coordinates on $\Sigma$.
Using \eq{kijn} and \eq{tfa}, we can calculate the extrinsic curvature $K_j^i$. Since the wormhole is symmetric by construction, the jump of the extrinsic curvature is simply
\bean
[K^i_j]=2K^i_j\,,
\eean
which gives the following relations
\bean
[ K^{\tau}_\tau]\eqn \frac{f(2\ddot r g+\dot r^2 g')+f'(1+\dot r^2 g)}{f\sqrt g\sqrt{1+\dot r^2 g}} \,,\label{kss1} \\
 \left[K_{\theta}^{\theta}\right]\eqn [K_{\phi}^\phi]=\frac{h'\sqrt{g(1+\dot r^2 g)}}{gh} \label{kss2} \,, \\
 \left[K\right]\eqn \left[K^{\tau}_\tau\right]+2  \left[K_{\theta}^{\theta}\right] \,.\label{kss3}
\eean

The surface stress tensor of a perfect fluid is given by
\bean
S^i_j=\left(\begin{array}{ccc}
-\sigma & 0 &0 \\
0& p & 0\\
0&0&p \end{array} \right) \,.\label{sij}
\eean
So $S=2p-\sigma$.
Using \eq{j3} and \meq{sij}, we find
\bean
\sigma\eqn -\frac{\phi}{8\pi}\left(\frac{\omega+1}{\omega}[K]-[K_\tau^\tau]\right)\,, \label{sr}\\
p\eqn  \frac{\phi}{8\pi}\left(\frac{\omega+1}{\omega}[K]-[K_\theta^\theta]\right) \,, \label{pr}
\eean
and thus
\bean
\left[K_\tau^\tau\right]\eqn \frac{8\pi}{(3+2\omega)\phi}\left[(2+\omega)\sigma+(2+2\omega)p\right] \,, \label{ik1}\\
\left[K_\theta^\theta\right]\eqn -\frac{8\pi}{(3+2\omega)\phi}(p+\sigma+\omega\sigma) \,.\label{ik2}
\eean

By solving \eq{kss2} for $\dot r^2$, one finds
\bean
\dot r^2=\frac{g h^2 \left[K_\theta^\theta\right]^2-h'^2}{gh'^2}\equiv V(r) \,. \label{rdv}
\eean

What we are interested in is a static wormhole solution, in which case the throat is located at a constant radius
\bean
r=a \,.
\eean
If the wormhole is stable at $r=a$, it requires
\bean
V(a)\eqn 0 \label{va}\,,\\
V'(a)\eqn 0 \label{vp} \,,\\
V''(a)&<&0 \label{v2p} \,.
\eean
Using \eq{rdv}, \eq{va} leads to
\bean
\left[K^\theta_\theta\right]\heq\frac{h'}{h\sqrt{g}} \,.
\eean
Here we have used $\heq$ to denote equations satisfied on the throat $r=a$, where \eqs{va} and \meq{vp} hold. Similarly, when evaluated on $r=a$, \eqs{kss1} and \meq{kss3} reduce to
\bean
\ktt&\heq&\frac{f'}{f\sqrt g} \,, \\
\left[K\right] &\heq &  \frac{f'}{f\sqrt g}+2\frac{h'}{h\sqrt{g}} \,.
\eean
Consequently, \eqs{sr} and \meq{pr} reduce to
\bean
\sigma&\heq& -\frac{\phi}{8\pi\sqrt{g}}\left[\frac{2h'}{h}+\frac{1}{\omega}\left(
\frac{f'}{f}+\frac{2h'}{h}\right)\right] \,, \label{sa}\\
p&\heq&
\frac{\phi}{8\pi\sqrt{g}}\left[\frac{h'}{h}+\frac{f'}{f}+\frac{1}{\omega}\left(
\frac{f'}{f}+\frac{2h'}{h}\right)\right] \label{pa}
\eean
and then
\bean
\sigma+p\heq\frac{\phi}{8\pi\sqrt{g}}\left[\frac{f'}{f}-\frac{h'}{h}\right]\,.
\eean

Now we calculate $V'(r)$. First, the conservation equation $\grad_a S^a_b=0$ yields
\bean
\dot\sigma =-\frac{\dot h}{h} (\sigma+p)\,,
\eean
where $\dot\sigma=\sigma'(r)\dot r$,$\dot h=h'(r)\dot r$. Thus, along the trajectory of the shell $r=r(\tau)$, we have
\bean
\sigma' =-\frac{h'}{h} (\sigma+p) \label{conh}\,.
\eean

Differentiating \eq{rdv} with respect to $\tau$, we have
\bean
V'(r)=2\ddot r \,.
\eean
Solving \eq{kss1} for $\ddot r$, we obtain
\bean
V'(r)=\frac{[K_\tau^\tau]f\sqrt g\sqrt{1+\dot r^2 g}-f'-\dot r^2 g f'-\dot r^2 f g'}{fg} \,. \label{vpr}
\eean

Now we  are ready to compute $V''(a)$ from \eq{vpr}.  Note that $\dot r^2=V(r)$ in \eq{vpr} and thus its derivative vanishes at $r=a$. So the key step is to compute $\ktt'$. Differentiating \eq{ik1} gives
\bean
\ktt'=\frac{8\pi}{(3+2\omega)\phi^2}\left[\phi((2+2\omega)p'+(2+\omega)\sigma')-\phi'((2+2\omega)p+(2+\omega)\sigma)
 \right] \,.  \label{kttp}
\eean
Now we insert \eq{conh} and assume \cite{tw1}
\bean
p'=\beta^2 \sigma'\,.
\eean
Then $\sigma'$ and $p'$ can be eliminated from \eq{kttp}. Here $\beta$ may be interpreted as the speed of sound for normal matter. Substituting  \eq{sa} and \eq{pa}, we obtain
\bean
\ktt'\hat =\frac{-(2+\omega+2\beta^2(1+\omega))hh'\phi f'+(2+\omega+2\beta^2(1+\omega))fh'^2\phi-(3+2\omega)h^2f'\phi'}{(3+2\omega)f\sqrt g h^2\phi} \,. \label{ktf}
\eean
Finally, by differentiating  \eq{vpr} and making use of \eq{ktf} and $\dot r=0$, we find
\bean
V''(a)&\heq& \frac{1}{(a^2-2\eta a)^2(2\omega+3)}\left(1-\frac{2\eta}{a}\right)^{-B}\left[ \left(2A^2\eta^2+4A(B-1)\eta^2+4A\eta a\right)(2\omega+3) \right.\non
&+&4(a+(B-1)\eta) (a+(B-A-1)\eta)(\omega+2+2(\omega+1)\beta^2) \left. \right] \,.\label{v2r}
\eean

\section{Constraints} \label{cons}
To find stable configurations, we need to consider all possible constraints.
According to \cite{bd2009}, $\phi_0>0$ yields
\bean
\omega<-2 \spa \textrm{or} \spa \omega>-3/2 \label{rome}\,.
\eean
As we show explicitly in Appendix \ref{app} that no solution exists for $\omega>-3/2$. Thus, we shall only consider $\omega<-2$ in the rest of this paper.

Since $\lambda$ is real, \eq{laxi} implies
\bean
\frac{-1+\sqrt{-3-2\omega}}{2+\omega}<\zeta<\frac{-1-\sqrt{-3-2\omega}}{2+\omega}\,.
 \label{cone}
\eean
It is also required \cite{bd2009}
\bean
B+1\geq 0 \,.
\eean
For $\omega<-2$, the solution is given by
\bean
2/\omega<\zeta<0 \spa \label{z1}
\eean

The energy conditions $\sigma\geq 0$ and $\sigma+p\geq 0$ at the throat $r=a$ require
\bean
a&\leq& -\frac{\eta}{\omega+1}\left[(B-1)(\omega+1)+\frac{A}{2}\right] \label{en1}\,,\\
a &\leq & \eta(A+1-B) \label{en2} \,.
\eean

To avoid singular behavior of the metric, the radius of the throat must satisfy
\bean
a>2\eta \,. \label{abt}
\eean
Finally, the junction condition \meq{j2} imposes the following constraint
\bean
a=-\frac{\eta}{2}\left[A+2B-2+\omega(A+B)\right] \,.\label{cfin}
\eean
A stable configuration must satisfy all the above constraints and the inequality \meq{v2p}. Next, we shall find out solutions satisfying all these constraints. For the two branches of $\omega$, we show in the appendix, that no solutions exist. Thus, we shall focus on $\omega<-2$ in the following derivation.

Without loss of generality, we shall take $\eta=1$ in the following calculation. By direct substitution, \eqs{en1}, \meq{en2} and \meq{cfin} can be written as functions of $\zeta$
\bean
a &\leq& f_\sigma(\zeta)\equiv \frac{\sqrt 2(1+2\omega+2\zeta(1+\omega))+2(1+\omega)\sqrt{2+2\zeta+\zeta^2(2+\omega)}}{2(1+\omega)\sqrt{2+2\zeta
+\zeta^2(2+\omega)}} \,,\nonumber \\
&& \label{fs}\\
a &\leq& f_{\sigma+p}(\zeta)\equiv \frac{2\sqrt 2+\sqrt 2 \zeta+\sqrt{2+2\zeta+\zeta^2(2+\omega)}}{\sqrt{2+2\zeta+\zeta^2(2+\omega)}} \,, \label{fsp}\\
a &=& f(\zeta)\equiv \frac{2\sqrt 2+\sqrt 2 \zeta(2+\omega)+\sqrt{2+2\zeta+\zeta^2(2+\omega)}}{\sqrt{2+2\zeta+\zeta^2(2+\omega)}} \,. \label{f}
\eean
 It is straightforward to show that
\bean
f_{\sigma+p}(\zeta)>f_\sigma(\zeta) \,.
\eean
Thus, inequality \meq{fsp} is implied by inequality \meq{fs}.

Moreover, we find
\bean
f_\sigma(\zeta)-f(\zeta)=\frac{\omega(1-\zeta-\omega \zeta)}{\sqrt 2(1+\omega)\sqrt{2+2\zeta+\zeta^2(2+\omega)}} \,. \label{fmf}
\eean
It then simplifies \eq{fs} as
\bean
\zeta>\frac{1}{1+\omega} \label{zfr}\,.
\eean
Note that
\bean
f'(\zeta)=\frac{3+2\omega}{\sqrt 2(2+2\zeta+(2+\omega)\zeta^2)^{3/2}}<0
\eean
for $\omega<-2$, which means $f(\zeta)$ decreases monotonically. Solving $f(\zeta)=2$, then \eq{abt} becomes
\bean
\zeta<\frac{-\omega-\sqrt 2\sqrt{3\omega+2\omega^2}}{2\omega+\omega^2} \,. \label{zoms}
\eean
It is not difficult to show that inequalities \meq{cone}, \meq{z1}, \meq{zfr},\meq{zoms} can be combined into one single inequality:
\bean
\zeta_1\equiv\frac{1}{1+\omega}<\zeta<\frac{-\omega-\sqrt 2\sqrt{3\omega+2\omega^2}}{2\omega+\omega^2}\equiv\zeta_2 \,.\label{main}
\eean
In this range, the WEC and all the constraints are satisfied. Therefore, we have derived the general solutions for the constraint equations in \cite{bd2009}.

\begin{figure}[htmb]
\centering \scalebox{0.6} {\includegraphics{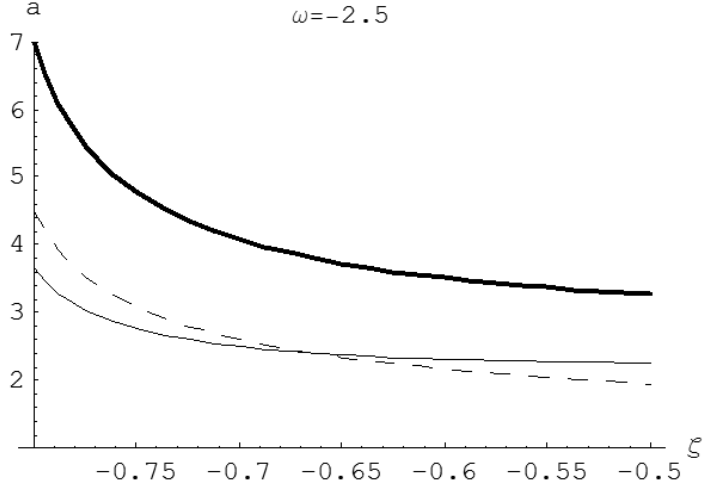}}
\caption{The region below the thick line satisfies $\sigma+p\geq 0$ and the region below the thin line satisfies $\sigma\geq 0$. The dashed line corresponds to \eq{cfin}. } \label{fig-axi}
\end{figure}

As an example, we consider $\omega=-2.5$ which has been chosen in \cite{bd2009}. \fig{fig-axi} depicts the region where constraints are satisfied.

\section{Stable solutions} \label{solution}

In section \ref{cons}, we have found all the solutions satisfying the WEC. Our task in this section is to check whether these solutions can be stable, i.e.,  whether $V''(a)<0$ can be satisfied. Note that from \eq{v2r}  the value of $V''(a)$ also depends on the parameter $\beta^2$. By substituting \eq{cfin} into \eq{v2r}, one can express $V''(a)$ as a function of $\zeta$
\bean
V''(\zeta)=\tilde V(\zeta)g(\zeta)\,,
\eean
where $g(\zeta)$ is positive in the allowed range of $\zeta$ and
\bean
\tilde V(\zeta)\eqn(-2\omega^2-2\beta^2\omega^2-\omega^3-2\beta^2\omega^3)\zeta^2+(2\omega+8\beta^2\omega+8 \beta^2\omega^2)\zeta\non
&+&(-6-\beta^2-3\omega-6\beta^2\omega) \,.
\eean
The stability condition $V''(\zeta)<0$ is equivalent to $\tilde V(\zeta)<0$
, i.e.,
\bean
2(1+\omega)(3-4\omega\zeta+\omega^2\zeta^2)\beta^2>-6-3\omega+2\omega\zeta-2\omega^2\zeta^2-\omega^3
\zeta^2\label{b22}
\eean
One can show \footnote{\eq{ome} is equivalent to $3-4\omega\zeta+\omega^2\zeta^2<0 $, which means
$\frac{3}{\omega}<\zeta<\frac {1}{\omega}$. One can check
$\frac{3}{\omega}<\zeta_1<\zeta_2<\frac{1}{\omega}$.
Therefore, in the range $\zeta_1<\zeta<\zeta_2$,
\eq{ome} always holds.}
\bean
(1+\omega)(3-4\omega\zeta+\omega^2\zeta^2)>0 \,. \label{ome}
\eean
\eq{b22} is equivalent to
\bean
\beta^2>f_\beta(\zeta)\equiv\frac{-6-3\omega+2\omega\zeta-2\omega^2\zeta^2-\omega^3
\zeta^2}{2(1+\omega)(3-4\omega\zeta+\omega^2\zeta^2)} \,.
\eean
Our goal is to find the minimum $f_\beta(\zeta)$ within the range $\zeta_1<\zeta<\zeta_2$ for each $\omega<-2$. We first consider
\bean
f_\beta'(\zeta)=\frac{\omega(3+2\omega)(-3+\omega^2\zeta^2)}{(1+\omega) (3-4\omega\zeta+\omega^2\zeta^2)^2} \,.
\eean
The condition
\bean
f_\beta'(\zeta)>0 \label{bep}
\eean
is equivalent to
\bean
-3+\omega^2\zeta^2<0 \,,
\eean
i.e.,
\bean
\zeta>\frac{\sqrt 3}{\omega} \,.\label{ie}
\eean
By solving
\bean
\frac{\sqrt 3}{\omega}<\zeta_1 \,.
\eean
we have
\bean
\omega<\frac{\sqrt 3}{1-\sqrt 3}\approx-2.366 \,.
\eean
In this range, \eq{ie} and then \eq{bep} always hold.
Consequently, for given $\omega$, the minimum required $\beta^2$ is given by
\bean
\beta_{min}^2(\omega)=f_\beta(\zeta=\zeta_1)=\frac{2+3\omega+2\omega^2}{-2-2\omega}\,,\label{bm1}
\eean
where $\zeta_1= \frac{1}{1+\omega}$.

For
\bean
-2.366<\omega<-2\,,
\eean
$f_\beta(\zeta)$ decreases in the range $\zeta_1<\zeta<\frac{\sqrt 3}{\omega}$ and increases in the range $\frac{\sqrt 3}{\omega}<\zeta<\zeta_2$. Hence,  The minimum value of $f_\beta(\zeta)$ is attained at $\zeta=\frac{\sqrt 3}{\omega}$. By substitution, we have the minimum as a function of $\omega$
\bean
\beta^2_{min}(\omega)=\frac{6-\sqrt 3+3\omega)}{2(-3+2\sqrt 3)(1+\omega)} \,. \label{bm2}
\eean

\begin{figure}[htmb]
\centering \scalebox{0.6} {\includegraphics{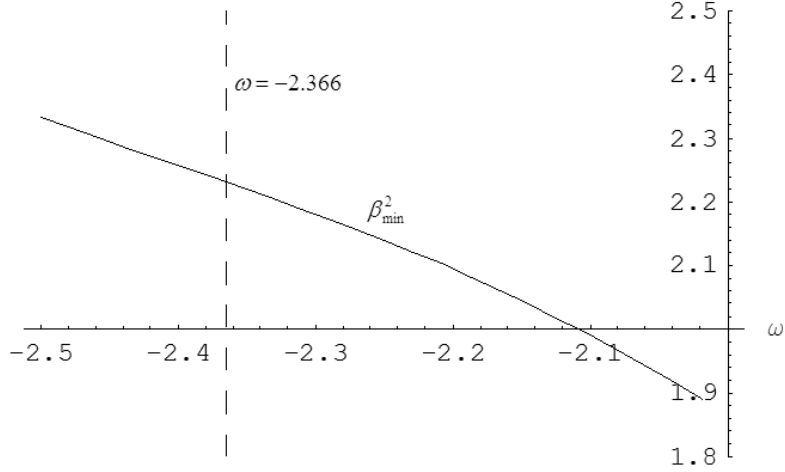}}
\caption{Plot of $\beta_{min}^2$ as a function of $\omega$. $\beta^2_{min}$ achieve a minimum at $\omega=-2$. } \label{fig-min}
\end{figure}

Combing \eqs{bm1} and \meq{bm2}, one has the function $\beta_{min}^2(\omega)$ defined in $-\infty<\omega<-2$. As depicted in \fig{fig-min}, we see $\beta_{min}^2$ changes smoothly across $\omega=-2.366$. Obviously, $\beta^2_{min}$ decreases monotonically with $\omega$ and attains its minimum value
 \bean
\beta^2_{0}=\frac{\sqrt 3}{2(-3+2\sqrt 3)}\approx 1.866
\eean
at $\omega=-2$. Hence no stable solutions exist for $\beta^2<\beta_0^2$.

\section{Conclusions} \label{conc}
We have reexamined the Brans-Dicke thin-shell wormhole model discussed in \cite{bd2009} and found all stable solutions satisfying the WEC.  The constraint equations listed in \cite{bd2009} have been solved analytically. These solutions require $\omega<-2$. We have also found  the exact parameter range of $\zeta$. For each $\omega$, the stability condition requires a minimum $\beta^2$ that decreases monotonically with $\omega$. Hence the minimum $\beta^2$ required for a stable solution is  $\beta_0^2=1.866$, which is attained at $\omega=-2$.

Our work shows that Brans-Dicke thin-shell wormholes can be supported by matter without violating the WEC and be stable against linear perturbation. To the best of our knowledge, this is the first example in four-dimensional spacetimes confirming  the existence of stable wormholes that do not violate the WEC. However, our calculation rules out any stable solutions for $\beta^2\leq 1$. This means the speed of sound at the throat of the wormhole must exceed the speed of light. Usually, such a solution is not acceptable because causality may be violated.
 However, whether superluminal behavior causes violation of causality has become a subtle and controversial issue in recent years (see e.g. \cite{c1}-\cite{c3}). We think the problem is still open and  the configurations with $\beta^2>1$ should not be ruled out.

\section*{Acknowledgements}
This research was supported by NSFC grants 10605006, 10975016, 10875012 and
by``the Fundamental Research Funds for the Central Universities''. We also thank two anonymous referees for their very constructive comments, which help us  significantly improve the quality of the manuscript.

\appendix
\section{Calculation for $\omega>-3/2$}  \label{app}
We shall show that no solutions exist in the range $\omega>-3/2$. Let us first consider the constraint $a>2$. It follows immediately from \eq{f} that
\bean
f(\zeta)>2  \,.\label{afz}
\eean
The roots of $f(\zeta)=2$ are
\bean
\zeta=\frac{-\omega\pm\sqrt 2\sqrt{3\omega+2\omega^2}}{2\omega+\omega^2}\,.
\eean
For $-3/2<\omega\leq 0$, the roots do not exist.
Note that $f'(\zeta)$ increases monotonically in this range. Thus, \eq{afz} cannot be fulfilled. So  the range of $\omega$ is restricted to $\omega>0$. In this case, \eq{afz} requires
\bean
\zeta>\frac{-\omega+\sqrt 2\sqrt{3\omega+2\omega^2}}{2\omega+\omega^2}>0\,.
\eean
Using \eq{fmf} , the constraint \meq{fs} is equivalent to
\bean
\zeta<\frac{1}{1+\omega}\,.
\eean
However, it is not hard to show that
\bean
\frac{-\omega+\sqrt 2\sqrt{3\omega+2\omega^2}}{2\omega+\omega^2}>\frac{1}{1+\omega}\,.
\eean
for any $\omega>0$. Therefore, no solutions satisfying the constraints can be found for $\omega>-3/2$.

\end{document}